\documentclass[AMA,STIX1COL]{WileyNJD-v2}
\usepackage{xcolor}
\usepackage{matlab-prettifier}
\usepackage{amsthm}
\usepackage{amsmath}
  \graphicspath{{./figs/}}
\usepackage[tight,footnotesize]{subfigure}

\articletype{Original Article}%

\begin{document}

\title{Orthogonal Rational Approximation of Transfer Functions for High-Frequency Circuits}

\author[1]{Andrew Ma}

\author[1]{Arif Ege Engin*}


\address[1]{\orgdiv{Department of Electrical and Computer Engineering}, \orgname{San Diego State University}, \orgaddress{\state{CA}, \country{U.S.A.}}}

\corres{*Arif Ege Engin, \email{aengin@sdsu.edu}}


\abstract[Summary]{Rational function approximations find applications in many areas including macromodeling of high-frequency circuits, model order reduction for controller design, interpolation and extrapolation of system responses, surrogate models for high-energy physics, and approximation of elementary mathematical functions. The unknown denominator polynomial of the model results in a non-linear problem, which can be replaced with successive solutions of linearized problems following the Sanathanan-Koerner (SK) iteration. An orthogonal basis can be obtained based on Arnoldi resulting in a stabilized SK iteration. We present an extension of the stabilized SK, called Orthogonal Rational Approximation (ORA), which ensures real polynomial coefficients and stable poles for realizability of electrical networks. We also introduce an efficient implementation of ORA for multi-port networks based on a block QR decomposition.}



\maketitle


\section{Introduction}
Rational function approximations find applications in many areas including macromodeling of high-frequency circuits \cite{gustavsen99}, model order reduction for controller design \cite{981109}, interpolation and extrapolation of system responses \cite{7577734}, surrogate models for high-energy physics \cite{AUSTIN2021107663}, and approximation of elementary mathematical functions \cite{hokanson2020multivariate}. 

\begin{figure*}[]
\centering
\includegraphics[width=0.9\columnwidth]{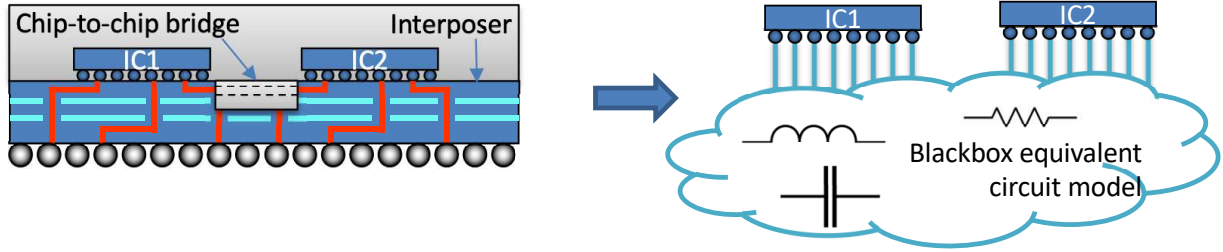}
\caption{Time-domain circuit simulation for signal and power integrity design of a system in package (SiP) requires a blackbox equivalent circuit model.}
\label{fig1}
\end{figure*}

Electromagnetic modeling of microelectronics packaging is a large field that benefits from accurate rational function approximations. Slow-down of Moore's law and economical concerns of yield are pushing the semiconductor industry towards heterogeneous integration, where multiple dies are interconnected through an advanced chip package, resulting in a System-in-Package (SiP). Heterogeneous integration allows a "more-than-Moore" approach that enables cutting-edge computing \cite{hi_amd, 9709842, 9621254, 7843744, 8847587}. In such advanced packaging, the interconnect parasitics can no longer be modeled using isolated circuit models available in closed-form \cite{engin05_2, 8382319} requiring blackbox models for time-domain circuit simulation of complex electromagnetic systems as shown in Fig.\ \ref{fig1}. Their description is however generally available as scattering parameters obtained from simulations or measurements. An intermediate step in generating an equivalent circuit model is a rational function approximation of this tabulated data \cite{9547288, Triverio+2021+275+310, NouriGadNakhlaAchar+2020+111+144, IoanCiuprinaSchilders+2020+145+200}. Available methods for rational transfer function approximation include the widely popular vector fitting \cite{gustavsen99}, Loewner framework \cite{MAYO2007634}, Sanathanan-Koerner (SK) iteration \cite{1105517}, and AAA (adaptive Antoulas-Anderson) \cite{doi:10.1137/16M1106122, gosea2020rational, 9512266, gosea2021algorithms}. 

Vector fitting is related to SK iteration \cite{1583747}. The partial fractions basis used in vector fitting and the iterative pole relocation has resulted in a robust method with wide range of successful applications. The implementation of SK iteration based on a monomial basis, on the other hand, becomes severely ill-conditioned due to two major reasons: the presence of Vandermonde matrices, and the weighting introduced by multiplication with the denominator from the previous iteration. One remedy to improve the conditioning of the method is to use orthogonal polynomials such as Chebyshev polynomials \cite{910554}, referred to as generalized SK iteration \cite{hokanson2018least}, and to properly scale the frequency variable. An orthogonal basis can also be generated on-the-fly at arbitrary frequency points using Vandermonde with Arnoldi \cite{doi:10.1137/19M130100X}. Similar approaches with orthogonal polynomial bases have also been studied for rational function approximation \cite{naivear, 781313,4026669,HOCHMAN2013337, AUSTIN2021107663}. A solution to address the second major source of ill-conditioning due to the weighting is introduced in the stabilized SK iteration \cite{hokanson2020multivariate}. 

In prior work, we have compared both polynomial and rational approximation methods for interpolation and least squares problems of scalar functions with no consideration on stability \cite{imaps21}. In this paper we address the approximation problem of multi-port networks with rational functions having real polynomial coefficients and stable poles. We extend the Arnoldi iteration method for calculating an orthogonal basis for the Vandermonde matrix \cite{doi:10.1137/19M130100X} to ensure real coefficients and integrate it in the stabilized SK iteration \cite{hokanson2020multivariate}. We extend this formulation to ensure stable poles, and introduce an efficient methodology for multi-port networks. The resulting orthogonal rational approximation (ORA) method is suitable for approximation of transfer functions with stable poles from tabulated data of multi-port network parameters.

\section{Orthogonal Rational Basis with Real Coefficients}
When performing a network parameter approximation using a rational function, the goal is a transfer function expressed as a ratio of two polynomials

\begin{equation}
    \frac{n(s)}{d(s)}=\frac{\sum_{i=0}^{n}a_i s^i}{\sum_{i=0}^{d}b_i s^i}.
\end{equation}

Assume that $m$ data samples $f_k$ are provided at frequency points $s_k=j\omega_k$. A least squares solution is desired that minimizes the residuals ${n(s_k)}/{d(s_k)}-f_k$ at the provided frequency points. This is however a nonlinear problem for the general case of unknown coefficients of the denominator polynomial. We will discuss the solution of this non-linear problem based on SK iteration in the next section. In this section, we introduce the orthogonalization of rational functions with real coefficients and assume for now that the denominator polynomial is known.

If $d(s)=1$, the problem reduces to a polynomial approximation. In this case, the Vandermonde matrix $A$ can be used to solve the coefficients $a$ of $n(s)$ that approximates the data in the least-squares sense
\begin{equation}
Aa\approx f.
\label{acf}
\end{equation}

The Vandermonde matrix is of size $m\times (n+1)$ and given as $A=[1\, s\, s^2\ldots s^n]$, where $s=(s_1, s_2,\ldots,s_m)^T$ is the vector of frequency points. An orthogonal basis for $A$ can be found with the $A=QR$ decomposition, which however results in an ill-conditioned least squares problem in (\ref{acf}) for high values of $n$. 

The Arnoldi iteration provides the same matrix $Q$ with orthogonal columns $[q_0\, q_1\, q_2\ldots q_n]$, with the advantage that the ill-conditioned matrix $A$ (and $R$) is never actually formed \cite{doi:10.1137/19M130100X}. It can be observed that the matrix $A$ is equivalent to
\begin{equation}
A=[q_0\, Sq_0\, S^2q_0\ldots S^n q_0],
\label{krylov}
\end{equation}
where the starting polynomial is $q_0=(1,1,\ldots,1)^T$ and $S=\text{diag}(s)$. The Arnoldi iteration is based on the application of Gram-Schmidt (GS) orthogonalization on the columns of $A$ in sequence. At step $i$, the vector $Sq_{i-1}$ is orthogonalized against the previous columns $[q_0\, q_1\, q_2\ldots q_{i-1}]$. Each column is chosen to have a norm of $\sqrt m$, to be consistent with the norm of $q_0$. The Arnoldi iteration also provides an $(n+1)\times n$ upper Hessenberg matrix $H$ that includes the coefficients used in orthogonalization process such that 
\begin{equation}
SQ_-=QH
\end{equation}
where $Q_-$ is obtained by removing the last column of $Q$. The polynomial coefficients $a$ are never calculated; instead, the least-squares problem is solved using transformed coefficients $c$ as: 
\begin{equation}
Qc\approx f.
\label{poly}
\end{equation} 

Generation of orthogonal polynomials for data fitting has been known for many decades (see e.g., \cite{10.2307/2098723}). More recently, the evaluation of such a fitted polynomial (at arbitrary frequency points) has been streamlined \cite{doi:10.1137/19M130100X} without explicitly using the three-term recurrence coefficients. The key insight is the use of $H$ to that purpose. For evaluating the polynomial at another set of frequencies $\hat{S}$, the same operations based on the previously obtained $H$ are applied to simply obtain the matrix $\hat{Q}$
 \begin{equation}
\hat{S}\hat{Q}_-=\hat{Q}H.
\label{qhat}
\end{equation}

The data at this new set of frequencies can then be obtained as $\hat{Q}c$.

For realizability of electrical networks, transfer functions with real polynomial coefficients are needed. To ensure real coefficients, one option is to enforce conjugate symmetry by fitting the complex conjugate responses on both sides of the frequency axis \cite{naivear, 781313}, which however unnecessarily uses a complex-valued Arnoldi iteration. We present an alternative method that uses real arithmetic throughout to ensure real coefficients by using an updated initial vector and frequency matrix. For the polynomial fitting example, the vector of coefficients $a$ in (\ref{acf}) can be enforced to be real by stacking the real and imaginary parts of $A=A'+jA''$ and $f=f'+jf''$ as
\begin{equation}
\left[
\begin{array}{c}
A'\\A''
\end{array}
\right]
a\approx 
\left[
\begin{array}{c}
f'\\f''
\end{array}
\right].
\label{stack}
\end{equation}

We can now observe that the stacked matrix is equivalent to
\begin{equation}
\left[
\begin{array}{c}
A'\\A''
\end{array}
\right]
=
[q\; Xq\; X^2q \ldots X^n q],
\label{krylov}
\end{equation}
where
 \begin{equation}
q=\left[
\begin{array}{cc}
q_0'\\
q_0''
\end{array}
\right],\;
X=\left[
\begin{array}{cc}
S'&-S'' \\
S''&S'
\end{array}
\right].
\label{skmatreal}
\end{equation} 

The obtained orthogonal basis would also be in stacked form as 
\begin{equation}
\left[
\begin{array}{c}
Q'\\Q''
\end{array}
\right],
\end{equation}
where $Q=Q'+jQ''$. For the usual case of a pure imaginary $s$ vector, $X$ becomes skew symmetric (it has zeros on its diagonal), and a Lanczos procedure equivalent to Arnoldi can be obtained \cite{Greif2009IterativeSO}. We use the Lanczos procedure for skew-symmetric matrices in this paper. Of particular interest are the zeroes of the orthogonal polynomial, which can be obtained in general from the comrade matrix \cite{Barnett1975ACM} or the state-space approach \cite{4026669} as the eigenvalues of the matrix 
\begin{equation}
H_n^T - h e c_{0:n-1}^T /c_n,
\label{poles}
\end{equation}
where $H_n$ is an $n\times n$ matrix obtained from $H$ by removing its last row, $h$ is the bottom right element of $H$, $e$ is a vector of $n-1$ zeros followed by a 1 as its last element, and $c_{0:n-1}$ is the coefficient vector except for the last element $c_n$. Since this matrix is real, the obtained zeros are either real or come in complex conjugate pairs.

As the next special case, assume that there is an arbitrary denominator polynomial $d(s)$, but its value is known at the $m$ frequency points given by the vector $d=1/[d(s_1), d(s_2),\dots, d(s_m)]^T$. Assume also that $d$ is normalized to have a norm of $\sqrt m$. For rational approximation, this case of known denominator would correspond to obtaining the numerator polynomial after the poles have been extracted. This results in the least squares problem of
 \begin{equation}
DAa\approx f
\end{equation} 
where $D$ is an $m\times m$ diagonal matrix given as $D=\text{diag}(d)$. Using a polynomial orthogonal basis $Q_p$ as discussed yields
 \begin{equation}
DQ_p\hat{c}\approx f,
\end{equation} 
which however can still become ill-conditioned due to the multiplication with $D$. The stabilization of SK iteration can be obtained by addressing this problem through generating an orthogonal basis for $DA$ instead \cite{hokanson2020multivariate}. For a rational function with real coefficients, this can simply be achieved by changing the initial vector in (\ref{skmatreal}) to 
\begin{equation}
q=\left[
\begin{array}{cc}
d'\\
d''
\end{array}
\right]
\end{equation} 
to calculate an orthogonal basis $Q=Q'+jQ''$ for the rational function and finding the least squares solution as usual from
 \begin{equation}
\left[
\begin{array}{c}
Q'\\Q''
\end{array}
\right]
c\approx 
\left[
\begin{array}{c}
f'\\f''
\end{array}
\right].
\label{rationala}
\end{equation}

At this point, if it is desired to calculate the numerator polynomial only, the straightforward option is to use $D^{-1}Qc$. However, this is only applicable at the original frequency points. A more powerful alternative is to use the $H$ matrix to obtain $\hat{Q}_p$ at arbitrary set of frequency points and calculating $\hat{Q}_p c$. Note that $\hat{Q}_p$ may not be orthogonal even at the original frequency points. 

In this section we have introduced how selecting the initial vector of the Arnoldi iteration from a known denominator polynomial allows us to obtain orthogonal rational functions with real coefficients. The algorithm is implemented in the Matlab function \texttt{numfit} as shown in Fig.\ \ref{rationalfitA}. Next, we integrate these results in the SK iteration to obtain the denominator polynomial as well. 

\begin{figure*}[]
\begin{lstlisting}[style=Matlab-editor]
function [H,Q,fit,myss,err] = numfit(den,s,n,f)
% den: denominator, s=jw, n: num degree, f: data
m = length(s);
Q = [real(den);imag(den)]*sqrt(m)/norm(den);
H = zeros(n+1,n);
for k = 1:n %Lanczos for skew-symmetric matrix
    q1 = Q(1:m,k);
    q2 = Q(m+1:2*m,k);
    q = [-imag(s).*q2; imag(s).*q1];
    if k>1 
        H(k-1,k) = -H(k,k-1);
        q = q - H(k-1,k)*Q(:,k-1); 
    end        
    H(k+1,k) = norm(q)/sqrt(m);
    Q = [Q q/H(k+1,k)];
end
if nargin == 4 %data is provided
    g = Q\[real(f); imag(f)]; %coeffs of numerator
    fit = Q*g;
    fit = fit(1:end/2,:)+ 1i*fit(end/2+1:end,:);
    err = norm(fit(:)-f(:));
    c = Q\[ones(m,1); zeros(m,1)]; %coeffs of denominator
    e=zeros(n,1);
    e(n)=1;
    g=g';
    A=H(1:n,1:n)'-H(n+1,n)*e*c(1:n)'/c(n+1);
    B=H(n+1,n)*e/c(n+1);
    C=g(:,1:end-1)-g(:,end)*c(1:n)'/c(n+1);
    D=g(:,end)/c(n+1);
    myss=ss(A,B,C,D); %state-space assuming improper rational function    
end
\end{lstlisting}
\caption{Matlab implementation of (\ref{rationala}) that fits the numerator with real coefficients from data on the imaginary axis and prescribed denominator.}
\label{rationalfitA}
\end{figure*}

\section{Orthogonal Rational Approximation}
A simple linearized version of the rational approximation problem \cite{https://doi.org/10.1007/s00211-018-0948-4, 6429401} can be formulated as
\begin{equation}
\underset{a, b}{\text{minimize}}  \sum_{k=1}^m \left| {n(s_k)} - f_k{d(s_k)}\right|^2, \text{ s.t. }\left|\left| \left[\begin{array}{c}
a\\
b
\end{array}\right]\right|\right|=1
\label{naive}
\end{equation}
whose solution can be obtained from a singular value decomposition (SVD). This naive method may not provide the correct solution due to the linearization. A well-known method to compensate for this inaccuracy is the SK iteration. 

The SK iteration can be started with the linearized least squares problem of (\ref{naive}) to obtain an initial solution for the denominator polynomial $d_{pre}(s)$. In order to approach the correct norm, the following problem is then solved to obtain updated $n(s)$ and $d(s)$ polynomials:
\begin{equation}
\underset{a, b}{\text{minimize}} \sum_{k=1}^m \left| \frac{n(s_k)}{d_{pre}(s_k)} - f_k\frac{{d(s_k)}}{d_{pre}(s_k)}\right|^2
\label{sk}
\end{equation}
with a suitable non-triviality constraint that we will discuss at the end of this section. This process can be continued iteratively until $d_{pre}(s)\approx d(s)$ so the correct norm is recovered if the algorithm converges. 

Using an orthogonal polynomial basis has been studied for rational function approximation before (see e.g., \cite{naivear, 781313,4026669,HOCHMAN2013337, AUSTIN2021107663}), where the coefficients of the orthogonal polynomial basis can be obtained using a three-term recurrence relation.   We emphasize that an orthogonal basis is needed for the rational functions of $n(s)/d_{pre}(s)$ and $d(s)/d_{pre}(s)$, and not merely the polynomials $n(s)$ and $d(s)$. The improvement in accuracy using orthogonal rational functions, rather than ortogonal polynomials can be dramatic as we will demonstrate in the numerical results. A vector fitting method based on an orthogonal basis of partial fractions is also available \cite{4214901}. Our method is not based on a partial fractions basis, therefore it is numerically different and we argue that it is simpler, especially in enforcing real coefficients for the polynomials. An advantage of our method is its flexibility to start the SK iteration with an arbitrary denominator polynomial (typically $d_{pre}(s)=1$ is selected) in addition to the usual selection of initial poles in vector fitting. This would for example allow to start the iteration with an initial choice of a denominator polynomial for numerical stability \cite{https://doi.org/10.1007/s00211-018-0948-4}. It is also possible to fit rational functions with a relative degree greater than 1.

The orthogonal rational approximation (ORA) is based on generating the orthogonal rational basis $Q_n$ for $n(s)/d_{pre}(s)$ and $Q_d$ for $d(s)/d_{pre}(s)$. If the degree of the numerator and denominator polynomials are equivalent, $Q_n=Q_d$; otherwise one can be obtained from the other by removing its last columns depending on the difference in degree. Assuming the diagonal matrix $F$ contains the data values as $F=\text{diag}(f_1, f_2,\ldots,f_m)$, the least squares problem can be expressed as 

\begin{equation}
\left[
\begin{array}{ccccc}
Q_n' & -(FQ_d)'\\
Q_n''  & -(FQ_d)''
\end{array}
\right]
\left[\begin{array}{c}
g\\
c
\end{array}
\right]
\approx 0.
\label{single}
\end{equation} 

At each step of the iteration, the roots of $d_{pre}$ are calculated and any unstable poles are flipped to obtain a set of stable poles $p_i$ similar to the process in vector fitting. The stable denominator polynomial is then calculated from $d_{pre}(s_k)=\prod_{i=1}^d{w(s_k-p_i)}$, using a roughly chosen weight $w$ to prevent overflow.

As for preventing the non-trivial solution, the straightforward choice is enforcing that the solution vector in  (\ref{single}) has a norm of 1. The solution is then obtained through the SVD of the matrix in (\ref{single}). 

An alternative is enforcing that condition on only the denominator coefficients $c$, and not having any constraints on $g$. In that case, the denominator polynomial with the coefficients vector $c$ can be calculated first, followed by calculating the numerator coefficients $g$ in a second step. Assume we obtain a QR decomposition of the matrix in (\ref{single}), where the submatrices of R are given as $R_{11}, R_{12}, R_{22}$. The least squares problem with this alternative constraint can then be expressed as

\begin{equation}
\left[
\begin{array}{ccccc}
R_{11} & R_{12}\\
0  & R_{22}
\end{array}
\right]
\left[\begin{array}{c}
g\\
c
\end{array}
\right]
\approx 0,\text{ s.t. }
||c||=1.
\end{equation} 

The solution for this homogeneous equation is given by the eigenvector corresponding to the smallest eigenvalue of $R_{22}^T R_{22}$ \cite{inkila2005homogeneous}. This is equivalent to calculating the final right singular vector of $R_{22}$ using SVD. 

\section{Multi-Port Networks}
For a multi-port network, a common-pole model can be obtained. Assume that $N$ elements of a network matrix will be approximated using the data set $F_1, F_2,\dots,F_N$. Since all $N$ rational functions will be using the same denominator polynomial, we can fit them simultaneously as
\begin{equation}
\left[
\begin{array}{ccccc}
Q_n' &0&\ldots&0& -(F_1Q_d)'\\
Q_n''  &0&\ldots&0& -(F_1Q_d)''\\
0&Q_n' &\ldots&0& -(F_2Q_d)'\\
0&Q_n''  &\ldots&0& -(F_2Q_d)''\\
\vdots&\ddots  &\ddots&\vdots& \vdots\\
0&0 &\ldots&Q_n'& -(F_N Q_d)'\\
0&0  &\ldots&Q_n''& -(F_N Q_d)''

\end{array}
\right]
\left[\begin{array}{c}
g_1\\
g_2\\
\vdots\\
g_N\\
c
\end{array}
\right]
\approx 0.
\end{equation} 

To obtain a fast method for multi-port networks, we can enforce $||c||=1$ as the non-triviality constraint and obtain the denominator first. This method is similar to the handling of multi-port networks in the fast implementation of the vector fitting algorithm \cite{4530747} and the parametric macromodeling approach \cite{8495287}. The least squares problem then reduces to
\begin{equation}
\left[
\begin{array}{c}
R_{22}^1\\
R_{22}^2\\
\vdots\\
R_{22}^N\\
\end{array}
\right]
c
\approx 0, \text{ s.t. }
||c||=1.
\label{oraa}
\end{equation} 

We improve the efficiency further by using a block QR decomposition for calculating the $R_{22}$ terms \cite{golub13}. Consider the matrix in (\ref{single}):
\begin{equation}
\left[
Q_1 \;B
\right]=
\left[
\begin{array}{cc}
Q_n' & -(FQ_d)'\\
Q_n''  & -(FQ_d)''
\end{array}
\right].
\end{equation} 

Since $Q_1$ is already orthogonal, the thin QR decomposition would be in the form of
\begin{equation}
\left[
Q_1 \; B
\right]=
\left[
Q_1 \; Q_2
\right]
\left[
\begin{array}{ccccc}
R_{11} & R_{12}\\
0  & R_{22}
\end{array}
\right].
\end{equation} 
Multiplying both sides from the left with $Q_1^T$ yields $Q_1^T B = mR_{12}$, where the factor of $m$ comes from our choice of having orthogonal columns with a norm of $\sqrt m$. We can now calculate $R_{22}$ from the QR decomposition of a smaller matrix $B-Q_1Q_1^T B/m=Q_2 R_{22}$. This is an additional advantage for ORA. The orthogonal rational functions not only improve the numerical conditioning, but also speed up the computation for multi-port networks. 

The implementation of denominator fitting in ORA is shown in Fig.\ \ref{ora}. Initially \texttt{denfit} can be called with a $1$ vector for the denominator. At each step, \texttt{denfit} is called with the denominator from the previous iteration.  

\begin{figure*}[]
\begin{lstlisting}[style=Matlab-editor]
function [den,poles] = denfit(den,s,n,d,f) 
%den: denominator, s=jw, n,d: degree of num and den, f: data
m = length(s);
[H,Q] = numfit(den,s,max(n,d));
Qd1 = Q(1:m,1:d+1);
Qd2 = Q(m+1:2*m,1:d+1);
Q = Q(:,1:n+1);
A = [];
for k=1:size(f,2)
    fr=real(f(:,k));
    fi=imag(f(:,k));
    Ar = -[fr.*Qd1-fi.*Qd2;fr.*Qd2+fi.*Qd1];
    Ar = Ar - Q*(Q.'*Ar)/m;
    [~,R] = qr(Ar,0);
    A = [A;R];
end
[~,~,V] = svd(A,0);
c = V(:,end);
poles = eig(H(1:d,1:d) - H(d+1,d)*(1/c(end))*c(1:end-1)*flip(eye(1,d)));
poles(find(real(poles)>0)) = poles(find(real(poles)>0))*-1; %pole flipping
den = 1./((prod((s.'-poles)./mean(abs(s)))).');
\end{lstlisting}
\caption{Matlab implementation of (\ref{oraa}) that updates the denominator.}
\label{ora}
\end{figure*}

\subsection{Integration in Circuit Solvers}
If needed, generating a transfer function in pole-residue form at this step is straightforward. Once the denominator polynomial is obtained, the poles of the rational function are readily available from the eigenvalues of (\ref{poles}). The residues can be fit in a similar way to the vector fitting residue identification process in a robust way. Alternatively, a direct state-space representation can be obtained from the coefficients of orthogonal polynomials for the case $(m<n)$ \cite{4026669}. This can be extended for the improper rational function $(m=n)$ case as

\begin{align}
A&=H_n^T - h e c^T_{0:n-1}/c_n\nonumber\\
B&=h e /c_n\nonumber\\
C&=(G_0 \; G_1\; \dots \; G_{n-1}) - G_n c^T_{0:n-1}/c_n\nonumber\\
D&=G_n/c_n
\end{align}
where $(G_0 \; G_1\; \dots \; G_{n})=(g_1 \; g_2\; \dots \; g_{N})^T$. After the denominator or the poles are extracted using iteratively calling \texttt{denfit}, a final call to \texttt{numfit} can provide this state-space model for improper rational functions. Note also how the zeroes of an orthogonal polynomial in (\ref{poles}) are calculated from the eigenvalues of $A$. 

\section{Numerical Examples}
\begin{figure}[]
\centering
\subfigure[]{
\includegraphics[width=0.48\columnwidth]{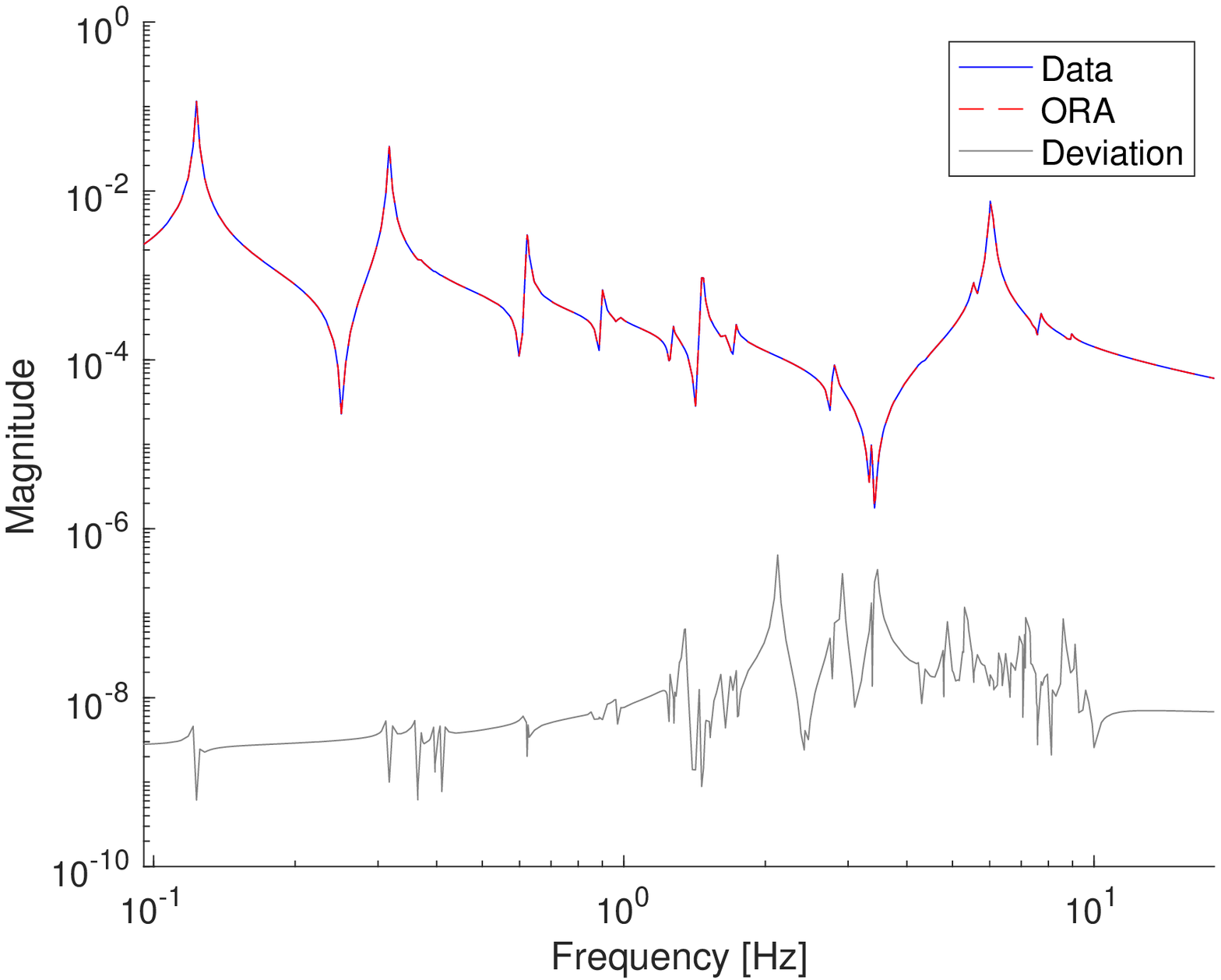}}
\subfigure[]{\includegraphics[width=0.48\columnwidth]{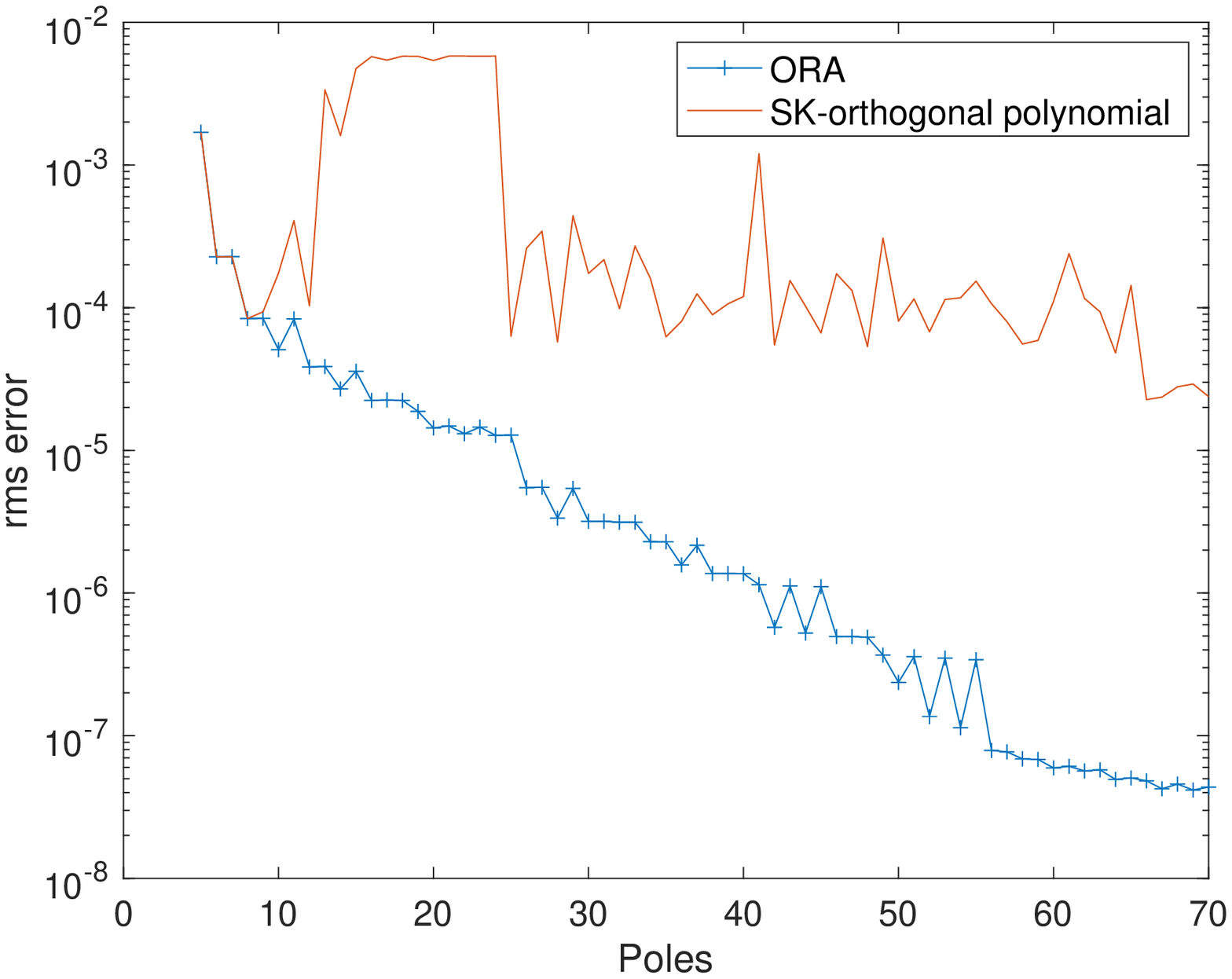}}\\
\caption{(a) Original and fitted data obtained using 70 poles in ORA for the ISS 1R module \cite{981109}. The error in magnitude is shown with light gray color. (b) rms error from the best fit among 20 iterations. SK with a merely orthogonal polynomial basis does not provide a well conditioned method.}
\label{rms}
\end{figure}

The first example we consider is the ISS 1R module \cite{981109}. The data is provided from a state-space model of order 260, so it does not include any noise. Fig. \ref{rms}(a) shows the original and fitted data for a model using 70 poles. We have assumed that the numerator and denominator polynomials have the same degree for the examples in this paper and have used 20 iterations.

Fig.\ \ref{rms}(b) shows a comparison of the rms error as a function of number of poles for the data in Fig.\ \ref{rms}. We selected the best fit among all the 20 iterations in this figure and started the SK iteration in ORA with $d(s)=1$. SK iteration based on merely an orthogonal polynomial basis becomes severely ill-conditioned due to the division by the denominator \cite{hokanson2020multivariate} as confirmed in Fig.\ \ref{rms}(b). This may be the reason for the poor accuracy observed in earlier implementations of the Arnoldi iteration for rational function approximation \cite{naivear} and why the method has so far not found widespread popularity. A well-conditioned method is recovered using the orthogonal rational function basis in ORA. 

\begin{figure}[]
\centering
\subfigure[]{
\includegraphics[width=0.48\columnwidth]{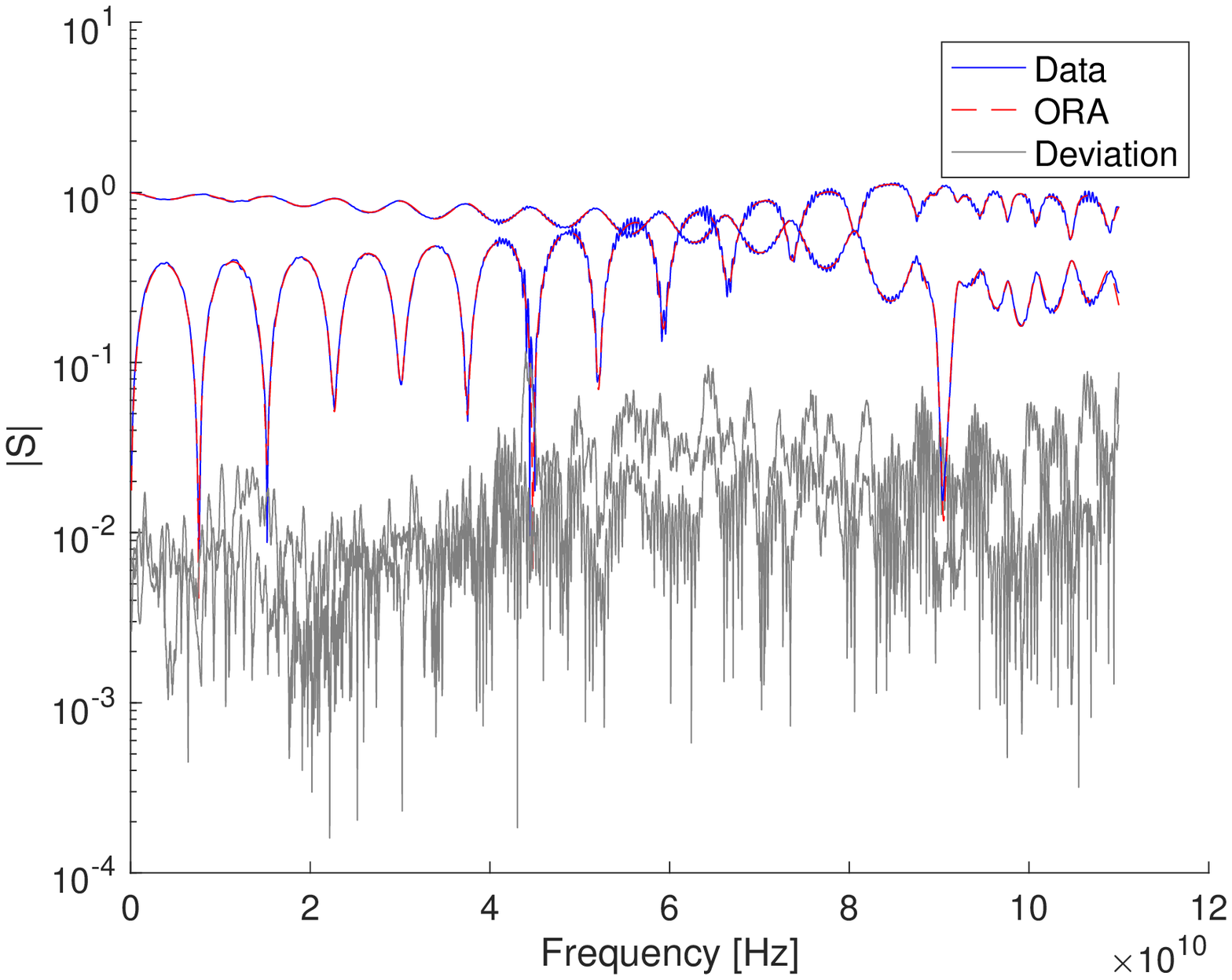}}
\subfigure[]{
\includegraphics[width=0.48\columnwidth]{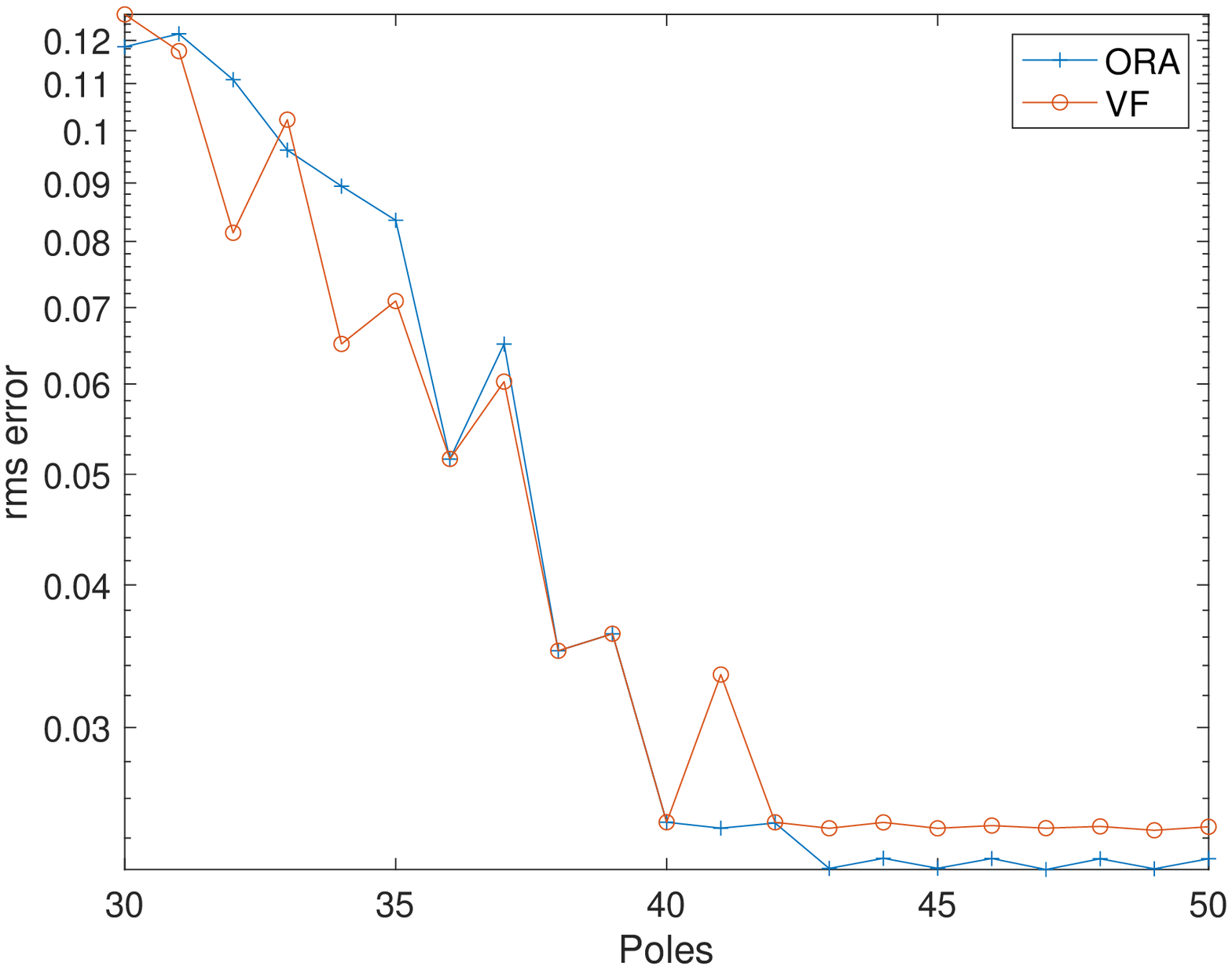}}
\caption{(a) ORA using 50 poles for the first row of the S-parameters of a stripline measured at 5001 frequency points. (b) rms error settles down to a lower level compared to vector fitting.}
\label{tline}
\end{figure}

The second example we consider is the noisy data of a stripline measured up to 110 GHz using 5001 frequency points on a vector network analyzer. The first row of the measured 2-port scattering parameters are approximated using ORA as shown in Fig.\ \ref{tline}(a). The proposed ORA method settles down to a lower residual error compared to the vector fitting method (vectfit3 implementation on Matlab) as shown in Fig.\ \ref{tline}(b) as the number of poles increases. 
 
\begin{figure}[]
\centering

\includegraphics[width=0.48\columnwidth]{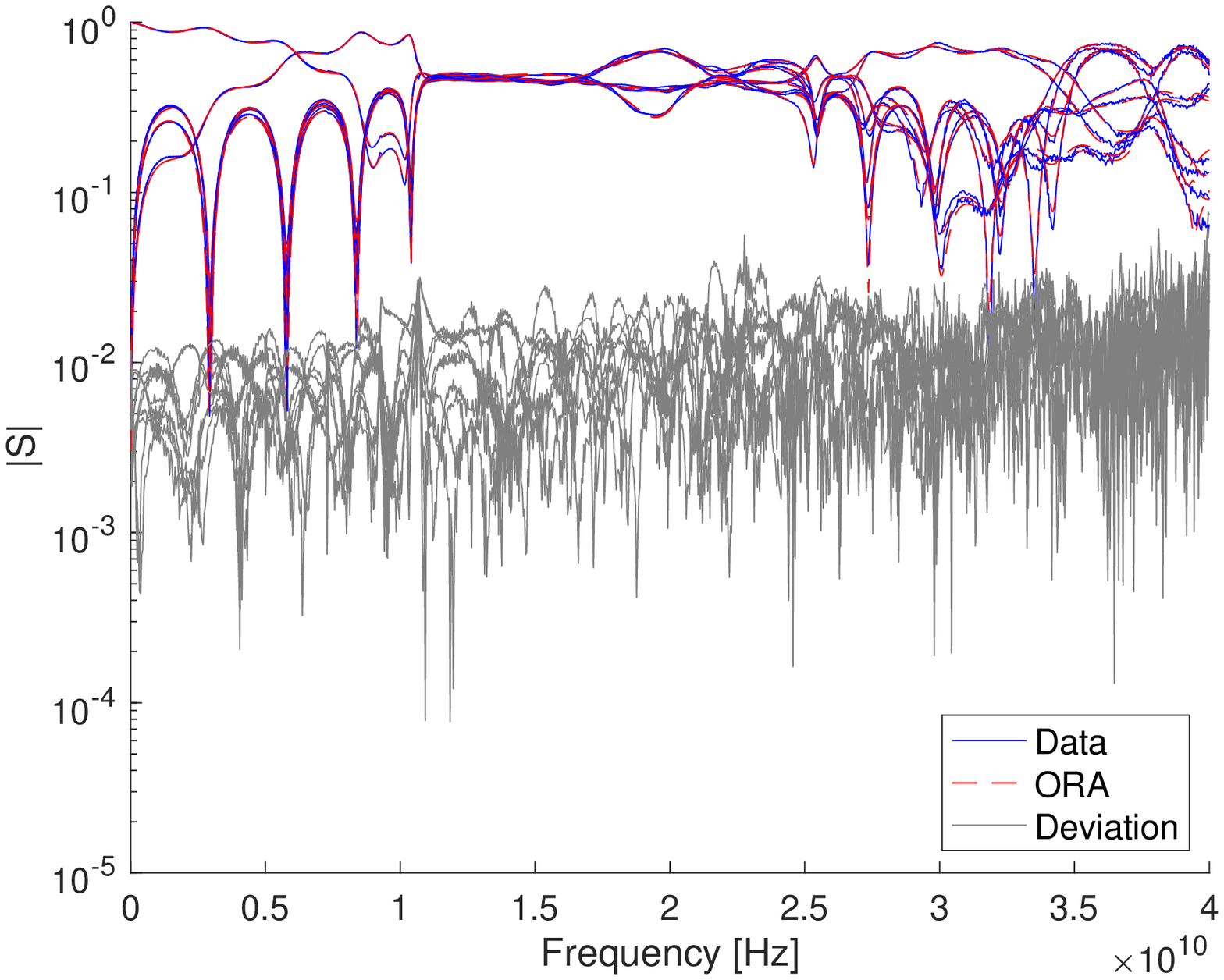}
\caption{Approximation of the upper triangular portion of the 4-port scattering matrix of a common-mode filter measured at 1001 frequency points. Good fit is obtained using 40 common poles.}
\label{cmf}
\end{figure}

The third example is a common-mode filter measured up to 40 GHz using 1001 frequency points. The upper triangular portion of the measured 4-port scattering matrix is approximated. Fig.\ \ref{cmf} shows good fit with a relatively flat residual using 40 poles.
 
\begin{figure}[]
\centering
\subfigure[]{
\includegraphics[width=0.48\columnwidth]{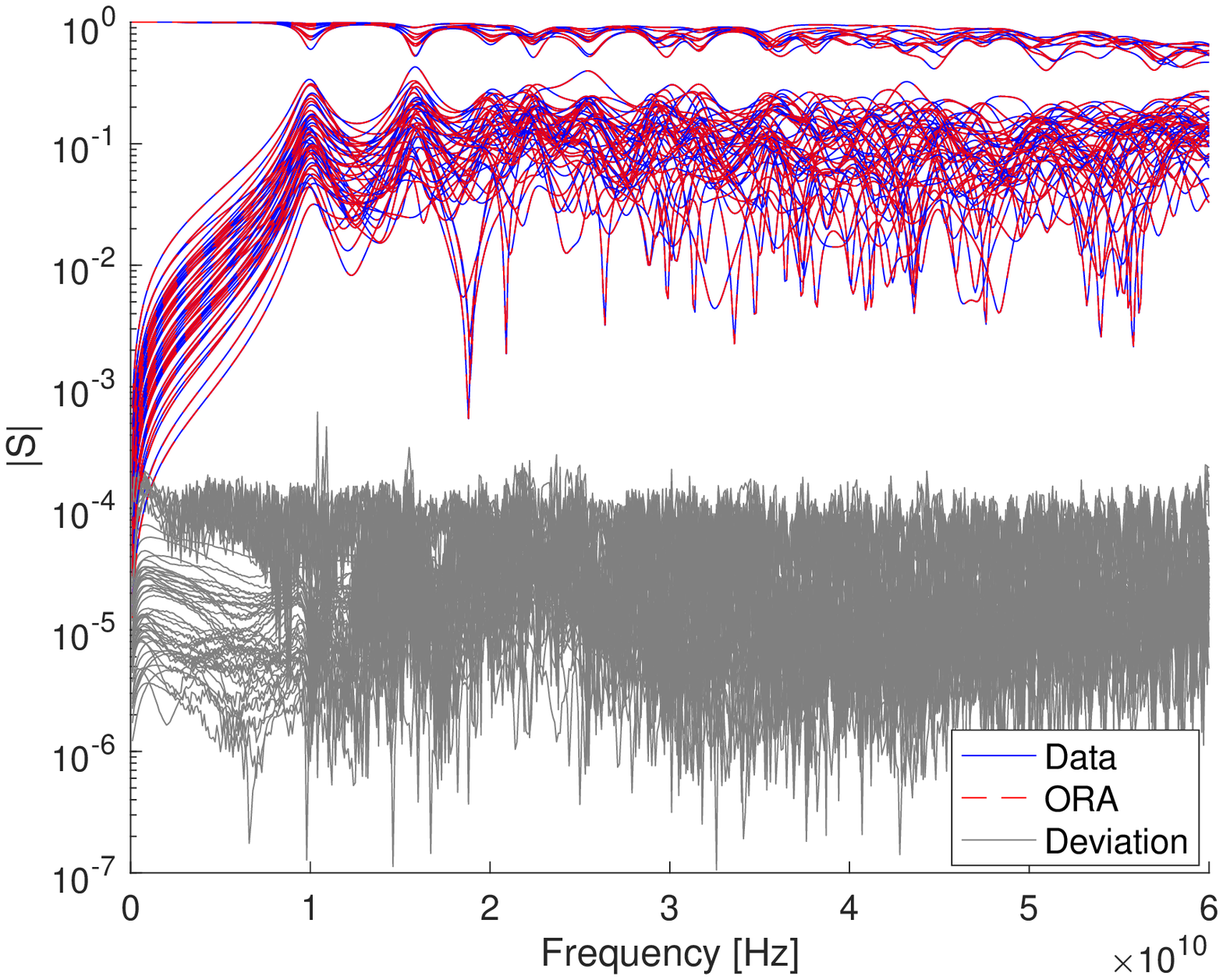}}
\subfigure[]{
\includegraphics[width=0.48\columnwidth]{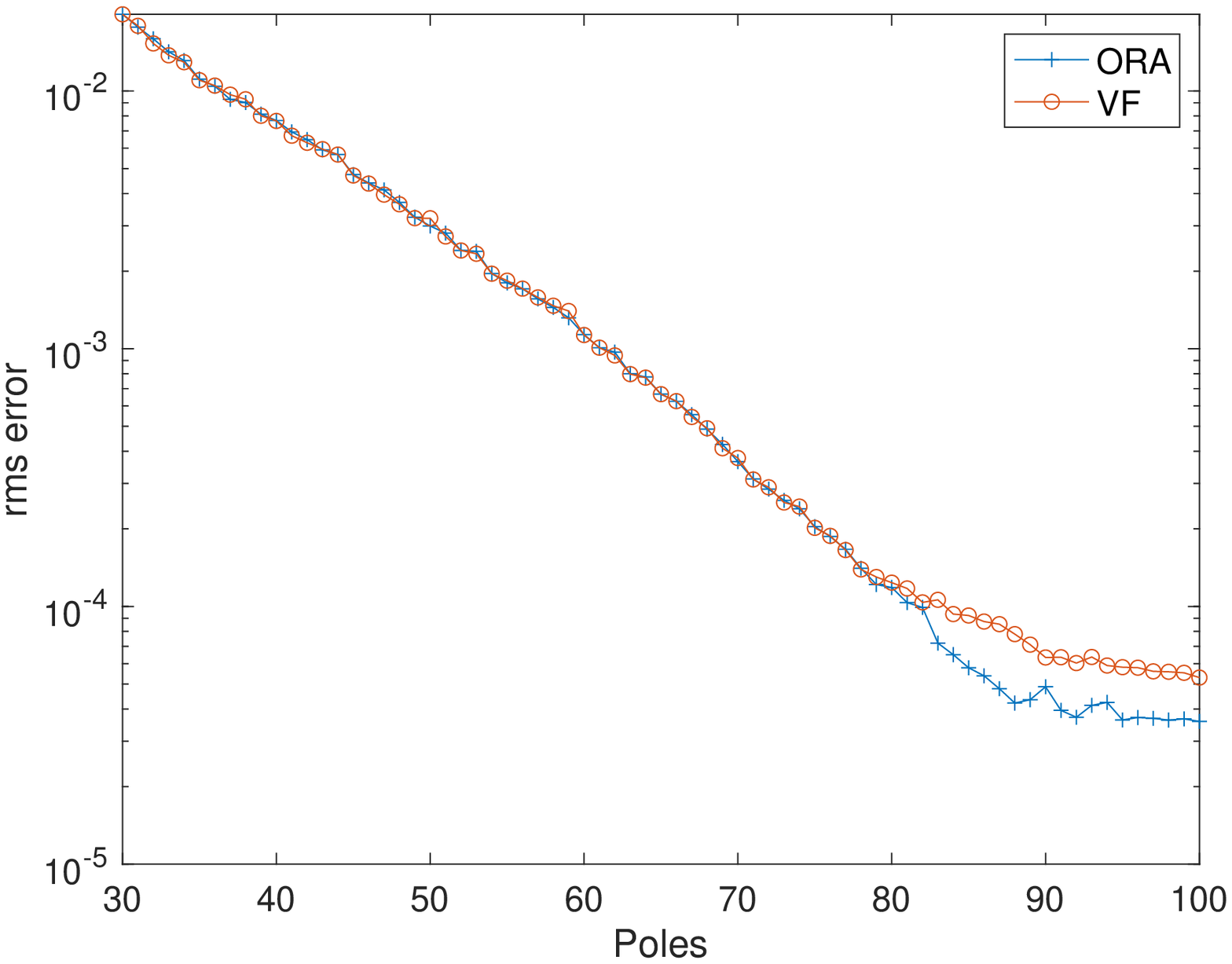}}
\caption{(a) ORA using 100 poles for the upper triangular portion of the S-parameters of a 10-port cavity resonator simulated at 600 frequency points. (b) ORA was able to find solutions with lower rms error compared to vector fitting as the model order increases.}
\label{sonnet}
\end{figure}

The final example is a cavity resonator simulated at ten ports using a full-wave simulator. The upper triangular portion of the scattering parameters (55 elements total) are approximated using ORA as shown in Fig.\ \ref{sonnet}(a). For this example, the proposed ORA method was also able to find solutions with lower residual error compared to the vector fitting method. The total run time for Fig.\ \ref{sonnet}(b) on a laptop with Intel i7 processor was also faster by a factor of approximately 2x: 166s for ORA vs.\ 336s for vector fitting. 

\section{Conclusions}
This paper introduced the Orthogonal Rational Approximation (ORA) method for rational function approximation. The method is an extension of the recently developed Vandermonde with Arnoldi and stabilized SK methods to ensure real polynomial coefficients and stable poles for realizability of the rational functions. The new method is also presented for multi-port networks and applied on rational function approximations of measured or simulated scattering parameters. For the considered examples, ORA showed a trend to find solutions with better accuracy compared to vector fitting as the model order is increased, where a 10-port model approximation also showed a speed up of approximately 2x. The presented method does not require an initial selection of poles and is well-conditioned due to the orthogonalization of rational functions in the SK iteration.
\bibliography{../bib/IEEEabrv,../bib/article,../bib/book,../bib/misc,}
\end{document}